\documentclass[aps,pre,reprint,floatfix,nofootinbib]{revtex4-1}
\pdfoutput=1
\usepackage{t1enc}
\usepackage{graphicx}
\usepackage{amssymb,amsfonts,amsmath}

\newcommand{\Rt}{\mathbf{R}_\text{T}}
\newcommand{\nm}{N_\text{m}}
\newcommand{\nss}{N_\text{ss}}
\newcommand{\nn}{n_\text{n}}
\newcommand{\nb}{n_\text{b}}

\newcommand{\nc}{n_\text{c}}
\newcommand{\uxx}{\varepsilon_{xx}}
\newcommand{\uyy}{\varepsilon_{yy}}
\newcommand{\nr}{n_\text{r}}
\newcommand{\tss}[1]{\mathbf{\tilde{t}}^{#1}}
\newcommand{\Tss}{\mathbf{\tilde{T}}}
\newcommand{\vare}{e}
\newcommand{\ea}{\vare_\text{a}}
\newcommand{\ex}{\vare_\text{x}}
\newcommand{\ey}{\vare_\text{y}}
\newcommand{\ena}{\vare_\text{na}}
\newcommand{\bs}{\boldsymbol}
\newcommand{\papertitle}{Selective buckling via states of self-stress in topological metamaterials}

\begin{document}
\title{\papertitle}
\author{Jayson Paulose}
\affiliation{Instituut-Lorentz, Universiteit Leiden, 2300 RA Leiden, The Netherlands}

\author{Anne S. Meeussen}
\affiliation{Instituut-Lorentz, Universiteit Leiden, 2300 RA Leiden, The Netherlands}

\author{Vincenzo Vitelli}
\email{vitelli@lorentz.leidenuniv.nl}
\affiliation{Instituut-Lorentz, Universiteit Leiden, 2300 RA Leiden, The Netherlands}

\maketitle

{\bfseries
States of self-stress, tensions and compressions of structural elements that result in zero net 
forces, play an important role in determining the load-bearing ability of 
structures ranging from bridges to metamaterials with tunable mechanical 
properties. We exploit a class of recently introduced states of self-stress 
analogous to topological quantum states to sculpt localized buckling regions 
in the interior of periodic cellular metamaterials. Although the topological 
states of self stress arise in the linear response of an idealized 
mechanical frame of harmonic springs connected by freely-hinged joints, they 
leave a distinct signature in the nonlinear buckling behaviour of a cellular 
material built out of elastic beams with rigid joints. The salient feature 
of these localized buckling regions is that they are indistinguishable from 
their surroundings as far as material parameters or connectivity of their 
constituent elements are concerned. Furthermore, they are robust against a wide 
range of structural perturbations. We demonstrate the effectiveness of this 
topological design through analytical and numerical calculations as well as 
buckling experiments performed on two- and three-dimensional metamaterials 
built out of stacked kagome lattices. 

}

Mechanical metamaterials are artificial structures whose unusual properties 
originate in the geometry of their constituents, rather than the specific material they are made of.
Such structures can be designed to achieve a specific linear elastic response, like 
auxetic (negative Poisson ratio)~\cite{Lakes1987} or pentamode (zero shear modulus)~\cite
{Kadic2012} elasticity. However, it is often their \emph{nonlinear} behaviour that 
is exploited to engineer highly responsive materials, whose properties
change drastically under applied stress or  
confinement~\cite{Shim2013,Babaee2013,Shan2014,Florijn2014,Driscoll2015}. Coordinated buckling 
of the building blocks of a metamaterial is a classic example of nonlinear 
behaviour that can be used to drive the auxetic response~\cite
{Bertoldi2009,Babaee2013}, modify the phononic properties~\cite{Shan2014} 
or generate 3D micro/nanomaterials from 2D templates~\cite{Xu2015}.

Buckling-like shape transitions in porous and cellular metamaterials 
involve large deformations from the initial shape, typically studied through finite element 
simulations. However, many aspects of the buckling behaviour can be 
successfully captured in an approximate description of the structure that is 
easier to analyze~\cite{Matsumoto2009, Kang2014,Shan2014,Florijn2014}. 
Here, we connect the mechanics of a cellular metamaterial, a foam-like 
structure made out of slender flexible elements~\cite{Gibson1989,Jang2013}, to that of a 
\emph{frame}---a simpler, idealized assembly of rigid beams connected by free 
hinges---with the same beam geometry. We exploit the \emph{linear} 
response of a recently introduced class of periodic frames~\cite{Kane2014}, inspired 
by topologically protected quantum materials, to induce a robust \emph
{nonlinear} buckling response in selected regions of two- and three-dimensional cellular 
metamaterials.

Frames, also known as trusses, are ubiquitous minimal models of mechanical
structures in civil engineering and materials science.
Their static response to an external load is obtained by balancing the forces
exerted on the freely-hinged nodes against internal stresses (tensions or
compressions) of the beams. However, a unique set of equilibrium stresses may
not always be found for any load~\cite{Pellegrino1993}. First, the structure may have loads that
cannot be carried because they excite hinge motions called \emph{zero modes}
that leave all beams unstressed. Second, the structure may support \emph{states
  of self-stress} -- combinations of tensions and compressions on the beams that
result in zero net forces on each hinge. An arbitrary linear combination of
states of self-stress can be added to an internal stress configuration without
disrupting static equilibrium, implying that degenerate stress solutions
exist for any load that can be carried.  The respective counts $\nm$ and $\nss$
of zero modes and states of
self-stress in a $d$-dimensional frame with $\nn$ nodes and $\nb$ beams are
related to each other by the generalized Maxwell
relation~\cite{Maxwell,Calladine1978}:
\begin{equation} \label{eqn_zmss}
  d\nn-\nb = \nm-\nss,
\end{equation}
As Eq.~\ref{eqn_zmss} shows, states of self-stress count the excess
constraints imposed by the beams on the $d\nn$ degrees of freedom provided by the
hinge positions.

States of self-stress play a special role in the mechanical response of
repetitive frames analyzed under periodic boundary
conditions~\cite{Guest2003}. Macroscopic stresses in such systems correspond to
boundary loads, which can only be balanced by states of
self-stress involving beams that cross the boundaries. If these states
span the entire system, boundary loads are borne by tensions and
compressions of beams throughout the structure. Conversely, by localizing states
of self-stress to a small portion of a frame, load-bearing ability is conferred
only to that region. Our approach consists of piling up states of self-stress in
a specific region of a repetitive frame so that the beams participating in these
states of self-stress are singled out to be compressed under a uniform load at
the boundary. In a cellular material with the same beam geometry, these beams
buckle when the compression exceeds their buckling threshold.

Although our strategy is of general applicability, it is particularly
suited to \emph{isostatic} lattices~\cite{Liu2010,Lubensky2015}, with $d\nn=\nb$
and no zero modes other than the $d$ rigid body motions under periodic boundary conditions. According to Eq.~\ref{eqn_zmss}, these lattices only have $d$ states of self-stress, insufficient to bear the $d(d+1)/2$ possible independent macroscopic stresses~\cite{Deshpande2001,Guest2003}. Additional states of self-stress can significantly
modify the load-bearing ability of these structures at the cusp of elastic stability.
Trivial states of self-stress can be created by locally adding extra beams 
between hinges. However, some isostatic 
periodic lattices can harbour topological states of self-stress  which owe 
their existence to the \emph{global} structure of the frame. These frames, 
introduced by Kane and Lubensky~\cite{Kane2014}, are characterized by a 
polarization $\Rt$ whose origin can be traced to topological 
invariants calculated from the geometry of the unit cell.
Domain walls between different topological polarizations as in Fig.~1a~\cite
{Kane2014,Chen2014}, and lattice defects~\cite{Paulose2015}, can harbour 
localized states of self-stress which can be used to drive localized buckling. Unlike their trivial counterparts, the existence of these topological states 
of self-stress cannot be discerned from a local count of the degrees of freedom or 
constraints in the region. An attractive feature for potential applications is their topological protection from perturbations of the lattice or changes in material 
parameters that do not close the acoustic gap of the 
structure~\cite{Prodan2009,Vitelli2012,Sun2012,Kane2014,Vitelli2014}.

In the remainder of this article, we use robust states of self-stress to design 
buckling regions in topological metamaterials composed of flexible beams 
rigidly connected to each other at junctions. As shown in Fig.~1, the states 
of self-stress are localized to a quasi-2D domain wall obtained by stacking 
multiple layers of a pattern based on a deformed kagome lattice~\cite{Kane2014}.   The domain 
wall separates regions with different orientations of the same repeating 
unit (and hence of the topological polarization $\Rt$ of the 
underlying frame). When the material is compressed uniaxially, beams 
participating in the topological states of self-stress will buckle out of 
their layers in the portion of the lattice highlighted in red, see 
Supplementary Movie 1. This region primed for buckling is indistinguishable from the remainder 
of the structure in terms of the density of beams per site and the 
characteristic beam slenderness. We verify using numerical calculations that 
the states of self-stress in the idealized frame influence the  beam 
compressions of the cellular structure in response to in-plane  compressions 
along the edges. The phenomenon survives even when the patterns on 
either side of the domain wall are nearly identical, reflecting the 
robustness of the topological design.

\begin{figure}
\centering
\includegraphics{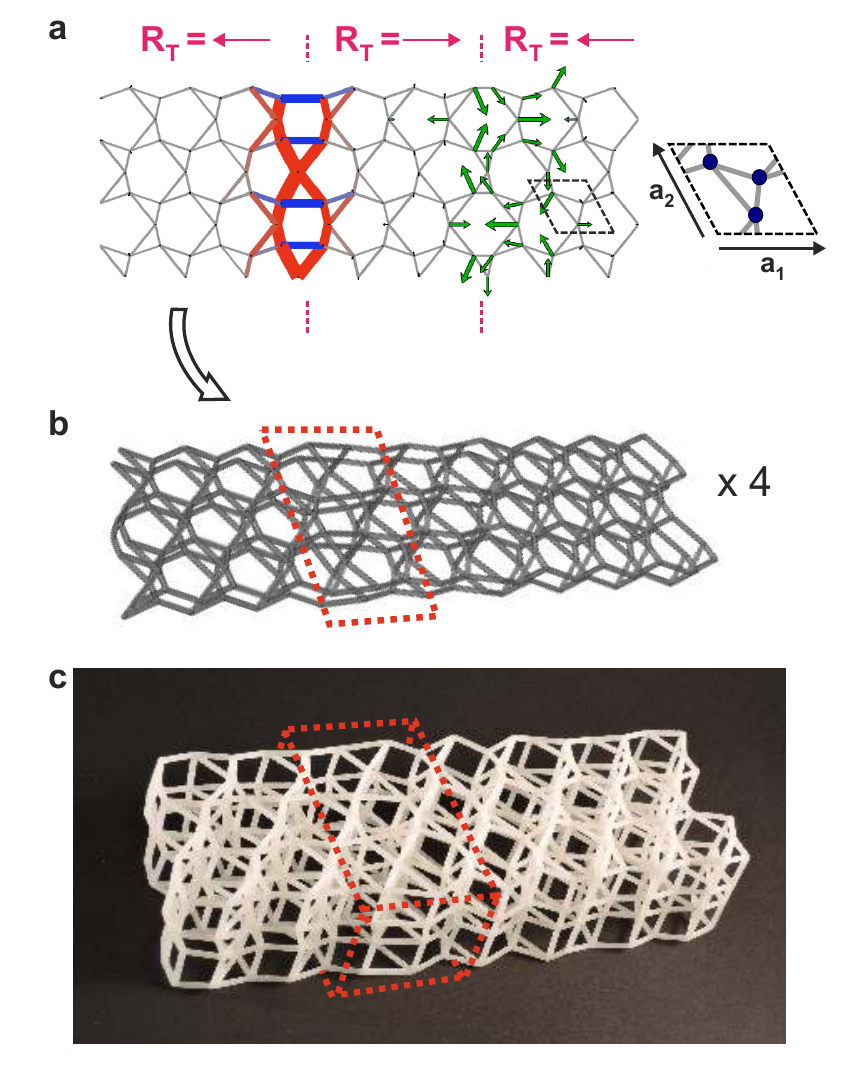}
    \caption{
        {\bfseries Topological buckling zone in a cellular metamaterial.}
        {\bfseries a,} Isostatic frame containing two domain walls that separate 
        regions built out of opposite orientations of the same repeating unit 
        (boxed; the zoom shows the three unique hinges in the unit cell as 
        discs). This unit cell carries a 
        polarization $\Rt=\mathbf{a}_1$ (solid arrow), and the 
        periodic frame displays topological edge modes~\cite{Kane2014}.
        The left domain wall harbours topological 
        states of self-stress, one of which is visualized by thickened beams 
        identifying equilibrium-maintaining tensions (red) and compressions 
        (blue) with magnitudes proportional to the thickness. The right domain wall 
        harbours zero modes, one of which is 
        visualized by green arrows showing relative displacements that do not 
        stretch or compress the beams. Modes were calculated using periodic 
        boundary conditions.
        {\bfseries b,} A 3D cellular metamaterial is obtained by stacking four 
        copies of the beam geometry in {\bfseries a}, and connecting 
        equivalent points with vertical beams to obtain a structure with 
        each interior points connecting 6 beams. The beams are rigidly 
        connected to each other at the nodes and have a finite thickness. Points are 
        perturbed by random amounts in the transverse direction and a small 
        offset is applied to each layer to break up straight lines of beams.
         A 3D-printed realization of 
         the design made of flexible plastic (see Materials and Methods)
         is shown in {\bfseries c}. The 
        sample has a unit cell size of 25 mm and beams with circular 
        cross-section of 2 mm diameter. The 
        stacking creates a pile-up of states of self-stress in a quasi-2D 
        region, highlighed by dotted lines.
    } \label{fig_intro}
\end{figure}

\section{Linear response of the frame}

The distinguishing feature of the isostatic periodic frame used in our design is the
existence of a topological characterization of the underlying phonon band 
structure~\cite{Kane2014}. If the only zero modes available to the structure are the $d$ rigid-body translations, its phonon spectrum has an acoustic gap; 
\emph{i.e.} all phonon modes have non-zero frequencies except for the translational modes at
zero wavevector. A gapped isostatic spectrum is  characterized by $d$ topological indices $n_i$, one for each primitive lattice vector $\mathbf{a}_i$ ($i \in\{1,2\}$ for the 2D lattice used in our design).
Although the phonon spectrum
can be changed by smoothly deforming the unit cell geometry, the indices themselves are
integer-valued and thus invariant to smooth perturbations. Their value can only be changed by
a deformation which closes the acoustic gap. Analogously to topological
electronic systems such as quantum Hall layers and topological
insulators which harbour protected edge
states~\cite{franz2013topological}, the existence of these invariants
guarantees the presence of zero modes or states of self-stress localized to
lattice edges or domain walls where the $n_i$ change value.

The count of topological mechanical states at an edge or domain wall is obtained
\emph{via} an electrostatic analogy.
The lattice vector $\Rt=\sum_i n_i\mathbf{a}_i$ (Fig.~1a) can be interpreted as a polarization of 
net degrees of freedom in the unit cell.
Just as Gauss's law yields the net charge enclosed
in a region from the flux of the \emph{electric} polarization through
its boundary, the net number of states of self-stress (minus the zero
modes) in an arbitrary portion of an isostatic lattice is given by the
flux of the \emph{topological} polarization through its
boundary~\cite{Kane2014}.  In Fig.~1a, the left domain wall has a net
outflux of polarization and harbours topological states of self-stress
(the right domain wall, with a polarization influx, harbours zero modes).  Although only one of
each mechanical state is shown, the number of states of self-stress
and zero modes is proportional to the length of the domain wall. Similar results can be 
obtained by using other isostatic lattices with an acoustic gap and a topological
polarization, such as the deformed square lattice in Ref.~\cite{Paulose2015}.

The linear response of a frame can be calculated from its equilibrium 
matrix $\mathbf{A}$, a linear operator which relates the beam tensions $\mathbf{t}$ (a vector 
with as many elements as the number of beams $\nb$) to the resultant forces on the 
nodes $\mathbf{p}$ (a vector with one element for each of the $2\nn$ degrees 
of freedom of the $\nn$ nodes) \emph{via} $\mathbf{A}\mathbf{t} = \mathbf{p}$. 
States of self-stress are vectors $\tss{q}$ which satisfy 
$\mathbf{A}\tss{q}=\mathbf{0}$, i.e. they are beam stresses which do not result in net forces on any 
nodes (the index $q$ identifies independent normalized states of self-stress 
that span the nullspace of $\mathbf{A}$). 
The same null vectors are also an orthogonal set of 
\emph{incompatible strains} of the structure, i.e. beam extensions that cannot be 
realized through any set of point displacements~\cite{Pellegrino1993}. 

Since we are interested in 
triggering buckling through uniform loads which do not pick out any 
specific region of the lattice, we focus on the response to affine 
strains, where affine beam extensions $\mathbf{\ea}$ are imposed 
geometrically by some uniform strain $\varepsilon_{ij}$ across the sample via
\begin{equation}
    {\ea}_\alpha = \hat{r}^\alpha_i \varepsilon_{ij} r^\alpha_j. \label{eqn_affineext}
\end{equation}
Here, $\alpha$ indexes the beams and $\mathbf{r}^\alpha$ is the end-to-end 
vector of beam $\alpha$. To attain equilibrium, the 
beams  take on additional non-affine extensions $\mathbf{\ena}$.
Under periodic boundary conditions, affine strains are  
balanced by loads across the system boundary rather than loads on specific 
nodes, which implies that the resultant beam tensions $\mathbf{t}_\text{a} = 
k(\mathbf{\ea}+\mathbf{\ena})$ must be constructed solely out of states of 
self-stress~\cite{Guest2003,Stenull2014} (we 
assume for simplicity that all beams have identical spring constant $k$). 
Therefore, $\mathbf{t}_\text{a} = \sum_q x_q \tss{q} \equiv 
\Tss\mathbf{x}$ where $\Tss = [\tss{1},\tss{2},...,\tss{\nss}]$ and the $x_q$ 
are the weights of the $\nss$ states of self-stress. These weights are 
determined by  requiring that the non-affine strains have zero overlap 
with the set of incompatible strains~\cite{Pellegrino1993} (the affine strains 
are automatically compatible with the affine node displacements):
\begin{equation}
    \Tss^\text{T}\mathbf{\ena} = 
    \Tss^\text{T}\left(\frac{1}{k}\Tss\mathbf{x}-\mathbf{\ea}\right) = \mathbf{0}, 
\end{equation}
which gives the solution 
\begin{align} 
    \mathbf{x} &= k \Tss^\text{T}\mathbf{\ea} \\
     \Rightarrow \mathbf{t}_\text{a} &= k\Tss\Tss^\text{T}\mathbf{\ea} =
    k \sum_q (\tss{q}\cdot\mathbf{\ea})\tss{q}. \label{eqn_sssum}
\end{align} 
Therefore, the beam tensions under an affine deformation are obtained by projecting the 
affine strains onto the space of states of self-stress.

Eq.~\ref{eqn_sssum} shows that the loading of beams under affine strains is 
completely determined by the states of self-stress. In a frame consisting of a 
single repeating unit cell, loads are borne uniformly across the structure. 
However, if the structure also has  additional states of 
self-stress with nonzero entries in $\tss{q}$ confined to a small region of the 
frame, it would locally enhance tensions and compressions in response to affine strains, 
provided the $\tss{q}$ have a nonzero overlap with the affine bond  
extensions imposed by the strain. Fig.~2 shows the
approximate states of self-stress\footnote{These approximate states of self-stress become exact when 
the separation between the two domain walls becomes large.} for the domain wall geometry of Fig.~1a 
which have a nonzero overlap with affine extensions $\mathbf\ex$ and $\mathbf 
\ey$ due to uniaxial strains $\varepsilon_{ij}=\delta_{ix}\delta_{jx}$ 
and $\varepsilon_{ij}=\delta_{iy}\delta_{jy}$ respectively.
 The system-spanning states of self-stress $\tss{a}$ and $\tss{b}$ in 
Figs.~2a--b respectively do not single out any particular region. Although they have a 
nonzero overlap with both $\mathbf\ex$ and $\mathbf\ey$, they do not provide 
significant stiffness to a uniaxial loading because a combined affine strain 
$\varepsilon_{ij} = \delta_{ix}\delta_{jx}+\beta \delta_{iy}\delta_{jy}$ exists with 
$\beta\approx-1.2$ such that the overlaps of the corresponding affine 
extensions with 
$\tss{a}$ and $\tss{b}$ are small (less than $10^{-5}$). Upon compression along 
one direction, say $y$, the frame can expand in the perpendicular 
direction to keep the tensions and compressions low in the majority of the 
sample. In contrast, the 
localized states of self-stress shown in Fig. 2c--d have a significant 
overlap with $\mathbf\ey$ but not 
$\mathbf\ex$. Therefore, according to Eq.~\ref{eqn_sssum}, a uniform  
compression applied to the lattice along the vertical direction, with the 
horizontal direction free to respond by expanding, will 
significantly stretch or compress only the beams participating in the localized states 
of self-stress. 

Whereas the topological polarization guarantees the presence of states of 
self-stress localized to the left domain wall, their overlap with one of the three 
independent affine strains is determined by the specific geometry of the 
hinges and bars. For the frame in Fig.~2,  
the states of self-stress visualized in Fig.~2c--d are crucial in triggering 
buckling response, which may be predicted to occur under compression along the 
$y$ direction (or extension along the $x$ direction, which would also lead to 
$y$-compression since the lattice has a positive Poisson ratio set by $\beta$). In other 
domain wall geometries, or other orientations of the domain wall relative to 
the lattice, the localized states of self-stress could have small overlap with 
affine strains, which would make them inconsequential to the buckling 
behaviour. Alternatively, states of self-stress which overlap significantly 
with shear deformations could also be realized which would enable buckling to 
be triggered via shear.

\begin{figure}
\centering
\includegraphics{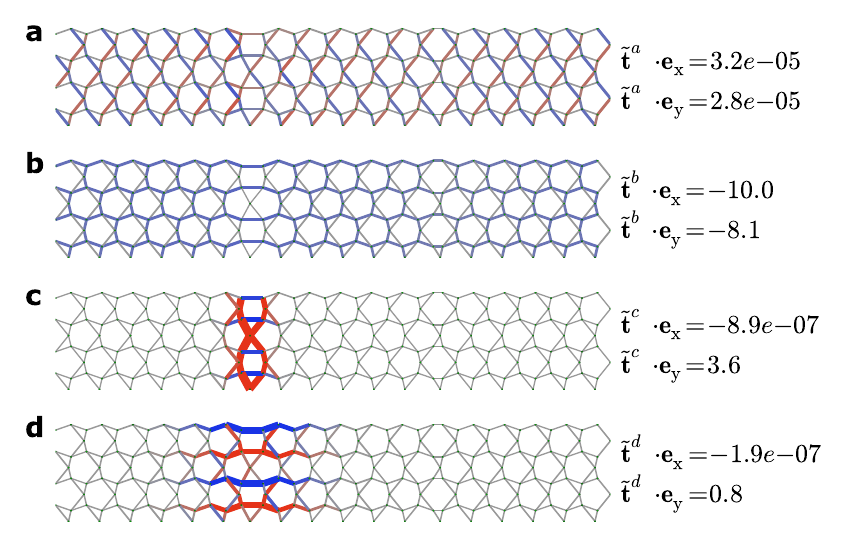}
    \caption{
        {\bfseries Extended and localized states of self-stress of the frame.} 
        {\bfseries a--d,} States of self-stress in the infinite periodic 
        frame obtained by tiling the design of Fig. 1a in both directions. 
        The states {\bfseries a, b} are 
        largely uniform over the structure, whereas {\bfseries c, d} are 
        localized to the left domain wall. The overlaps of each state of 
        self-stress with the affine strains $\mathbf\ex$ and $\mathbf\ey$ 
        (see text) are also shown. Only states of self-stress with significant 
        overlaps are shown; all other states of self-stress in the structure 
        have $\tss{q}\cdot\mathbf\vare_{\{x,y\}} < 10^{-5}$.
    } \label{fig_linear}
\end{figure}

\section{Buckling in topological cellular metamaterials}

\begin{figure}
\centering
\includegraphics{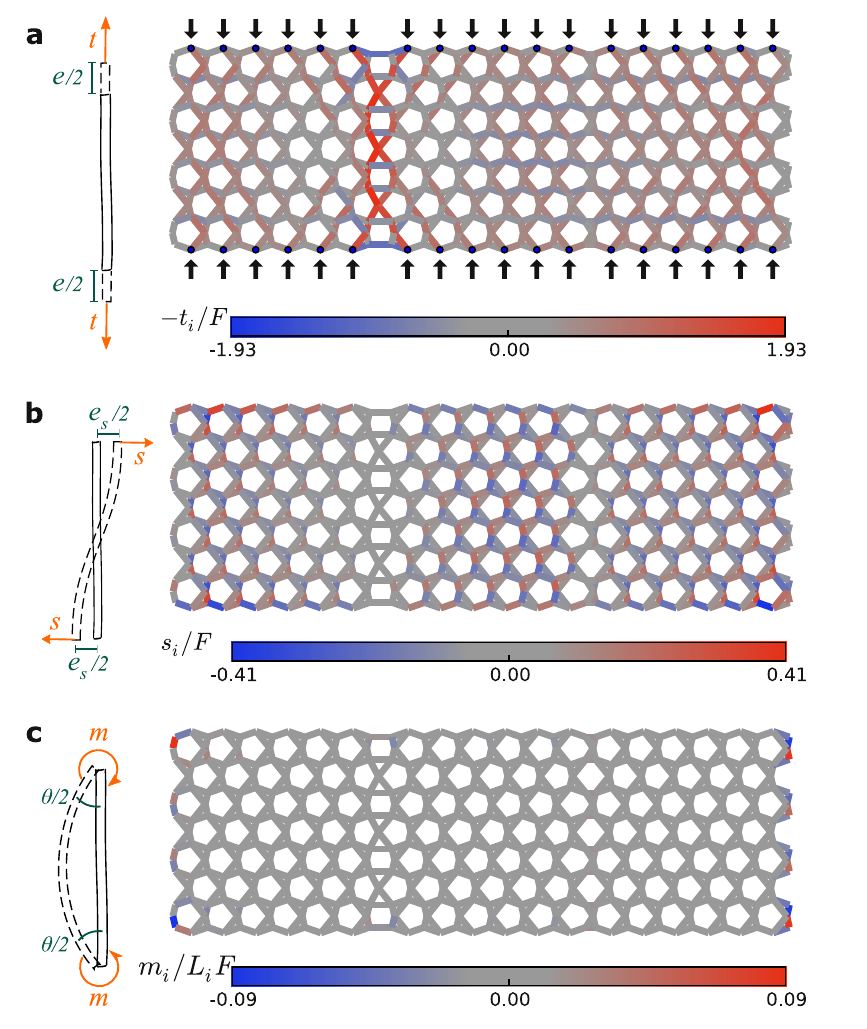}
    \caption{
        {\bfseries Stretching, shear and bending contributions to the linear 
        in-plane response of the cellular metamaterial.}
        Response of a planar cellular structure,
        related to the lattice in Fig.~2 but with free edges, 
        subject to a vertical compressive force $F$ 
        (solid arrows) at each point highlighted along the top and bottom 
        edges. The structure is modeled as a network of flexible beams 
        connected by rigid joints at the nodes, and with each beam providing 
        torsional stiffness in addition to axial stiffness. The beams are 
        coloured according to:
        {\bfseries a,} axial compression; {\bfseries b} shear load; 
        {\bfseries c,} bending moment.
    } \label{fig_cellular}
\end{figure}

The cellular metamaterial differs from the ideal frame in two important ways. 
First, real structures terminate at a boundary, and loads on the boundary are no 
longer equilibrated by states of self-stress, but rather by tension 
configurations that are in equilibrium with forces on the boundary 
nodes~\cite{Pellegrino1993}. However, these tension states are closely related 
to system-traversing states of self-stress in the periodic case, with the 
forces on the boundary nodes in the finite system playing the role of the 
tensions exerted by the boundary-crossing beams in the periodic system. 
Therefore, the states of self-stress also provide information about the load-bearing regions in 
the finite structure away from the boundary.

In addition to having boundaries, the cellular block probed in Fig.~1 also 
departs from the limit of an ideal frame, as it is made of flexible beams 
rigidly connected at the nodes and can support external loads through shear and
bending of the beams in addition to axial stretching or
compression. Nevertheless, the states of self-stress and tension states of the
corresponding frame (with the same beam geometry) determine the relative
importance of bending to stretching in the load-bearing ability of the cellular
structure~\cite{Deshpande2001}. 
To verify that the localized states of 
self-stress in the underlying frame influence the response of the finite cellular structure, we 
numerically calculate the in-plane response of each layer treated as an independent 2D cellular structure 
with loading confined to the 2D plane. Each beam provides not just axial 
tension/compression resistance but also resistance to shear and bending. A 
beam of length $L$, cross-sectional area $A$ and area moment of inertia $I$ resists 
(i) axial extensions $\vare$ with a tension $t = (EA/L)\vare$, (ii) 
transverse deformations $\vare_s$ with a restoring shear force $s = 
12(EI/L^3)\vare_s$, and (iii) angular deflections of the end nodes 
$\theta$ with restoring moment $m = (EI/L)\theta$ (Fig.~3).  Since $A \sim 
w^2$ and $I \sim w^4$ for a beam of width $w$, the relative 
contribution of the bending, shear and torsional components  
to the total stiffness is set by the aspect ratio $w/L$ of each beam; the frame 
limit with no bending or shear stiffness is recovered when $w/L \to 0$.  
The physical sample has beams with aspect ratios in the range $0.1 \lesssim w/L \lesssim 0.2$, indicating a small but appreciable contribution of shear and torsion to the response. 

The 2D linear response of such a structure is  
calculated by augmenting the equilibrium matrix $\mathbf{A}$ to include 
an additional degree of freedom (a rigid rotation angle) at each node, and two 
additional restoring forces (shear and torsion) for each beam, see Materials and Methods.  The resulting equilibrium matrix is of size $3\nn 
\times 3\nb$, which implies that cellular solids based on isostatic frames 
with $\nb \approx d\nn$ are severely overconstrained for $d = 2,3$, and 
structurally stable even with free boundaries. In particular, they can support 
any loads exerted on the boundary nodes as long as the net forces and torques 
about the center of mass are zero. 
Once the equilibrium matrix is constructed, its singular value decomposition 
can be used to completely determine all stresses and torques experienced by 
the beams in response to forces specified on the boundary 
nodes~\cite{Pellegrino1993}. Fig.~3 shows the linear response of each plane of the cellular 
pattern corresponding to the domain wall geometry of Fig.~2 under uniform compressive 
force applied to the nodes on the top and bottom edges. The beams 
participating in the states of self-stress at the left domain wall are singled 
out by their high compression, whereas the rest of the structure primarily 
supports the boundary load through shear rather than compression or bending.
Remarkably, the unique compression-dominated response of the left domain wall 
(originating from the topological invariant in the idealized isostatic frame with 
freely-hinged joints) survives 
in cellular structures away from the isostatic condition.

We expect a similar localized compression-dominated response in each 
layer of the stacked structure (Fig. 1b--c).
The enhanced compressions along the left domain wall trigger buckling when 
the compression exceeds the Euler beam buckling threshold, $t_i < -cEI/L_i^2$,
where $c$ is a positive numerical
factor determined by the clamping conditions as well as cooperative buckling
effects. Buckling is signified by a 
loss of ability to bear axial loads, as the beam releases its compression 
by bending out of plane. Upon compressing the 3D sample between two plates as 
shown in Fig.~4a, we see a significant out-of-plane deflection for beams along the 
left domain wall (Fig.~4b--c), consistent with buckling of the maximally 
stressed beams in Fig.~3a. The deflection in different layers is coordinated by the 
vertical beams connecting equivalent points, so that beams within the same 
column buckle either upwards (column 1) or downwards (column 2) to produce a distinctive 
visual signature when viewed along the compression axis. Beams 
connecting different planes in the stacked pattern create additional states 
of self-stress that traverse the sample vertically, but these do not single 
out any region of the material, and do not couple to the specific in-plane 
loading of Fig.~4a.

\begin{figure*}
  \centering
    \includegraphics{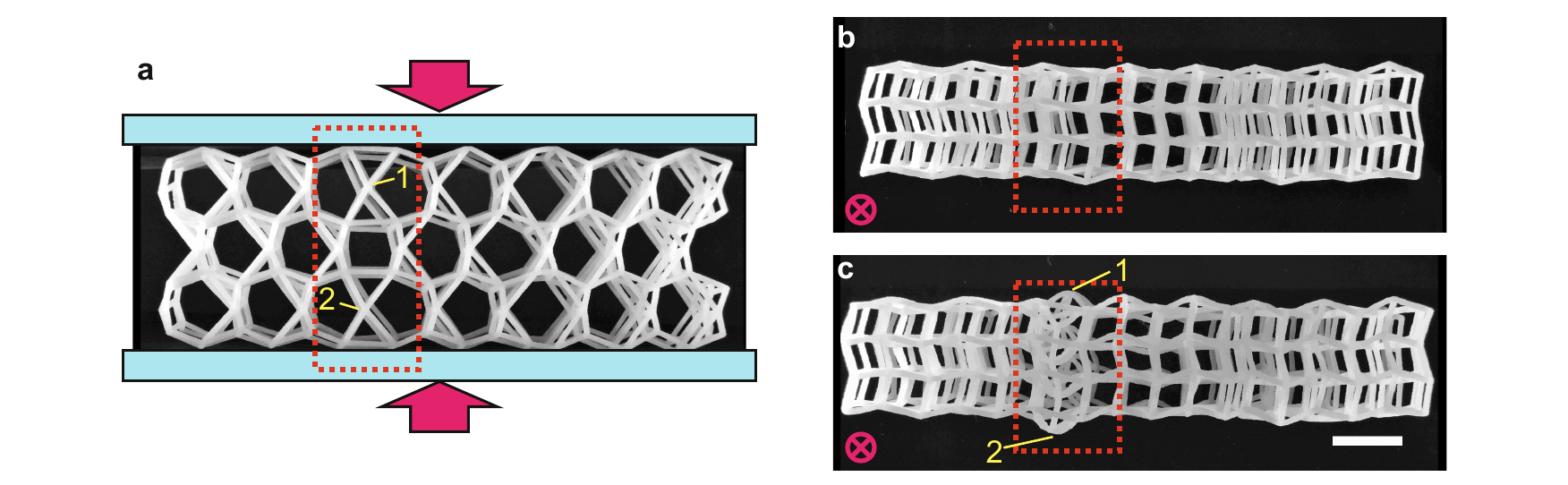}
    \caption{
        {\bfseries Buckling in the 3D topological cellular metamaterial.} 
        {\bfseries a,} Top view of the 3D sample whose construction was 
        outlined in Fig.~1, showing the compression 
        applied by confining the sample between transparent plates in contact with the 
        front and back surfaces. The buckling zone is highlighted in red as in 
        Fig.~1. Two vertical columns within this zone are labelled in yellow. 
        Magenta arrows show the compression direction.
        {\bfseries b--c,} View along the compression axis at compressions of 0 
        and 20\% respectively. The beams in the region with states of 
        self-stress have buckled in the vertical direction, whereas other 
        beams have largely deformed within their stacking planes. Scale bar, 
        25 mm.
    } \label{fig_3d}
\end{figure*}

We emphasize that having as many states of self-stress as there are 
unit cells along the domain wall is crucial for the buckling to occur 
throughout the domain wall. When a beam buckles, its contribution to the 
constraints of distances between nodes essentially disappears, reducing the 
number of load-bearing configurations by one. If there were only a single 
localized
state of self-stress in the system, the buckling of a single beam would eliminate 
this state, and the compressions on the other beams would be relaxed,  
preventing further buckling events. However, the presence of multiple states 
of self-stress, guaranteed in this case by the topological origin of the 
states, allows many buckling events along the domain wall. In the SI Text, we use an adaptive simulation that sequentially removes highly compressed beams to show 
that each repeating unit along the $y$ direction
experiences a unique buckling event even when the loss of constraints due to 
other buckling events is taken into account  (Supplementary Figure~S1).

\section{Robustness of the buckling region}
Finally, we show that the robustness of the topological states of 
self-stress predicted within linear elastic theory carries over to the 
buckling response in the non-linear regime. There is a wide range of distortions
of the deformed kagome unit cell which do not close the acoustic gap, leaving the topological
invariants $n_i$ and hence the polarization $\Rt$ unchanged~\cite{Kane2014}. These distortions may
bring the unit cell arbitrarily close to the regular kagome lattice with equilateral triangles of beams (for which the
gap is closed and the polarization is undefined). The unit cell of the two dimensional 
lattice shown in Fig.~5a is minimally distorted away from the regular kagome 
lattice, but this barely noticeable 
distortion (inset of Fig. 5a) is sufficient to induce the same topological 
polarization $\Rt$ in the underlying frame as before. As a result, the domain wall on 
the left (characterized by a net outflux of polarization) localizes states 
of self-stress (shown in Supplementary Figure~S2) even though 
the unit cells are nearly identical on either side. 

For ease of visualization, we tested this design in a 2D prototype cellular 
metamaterial, obtained by 
laser-cutting voids in a 1.5 cm thick slab of polyethylene foam (see Materials and Methods),
leaving 
behind beams of width 1--2 mm and lengths 10--12 mm (Fig.~5a). The aspect ratio of the beams is comparable to that of the 3D sample.  
Since the slab thickness is much larger than the beam width, 
deformations are essentially planar and uniform through the sample 
thickness, and can be captured by an overhead image of the selectively 
illuminated top surface. The sample was confined between rigid acrylic plates in free contact with 
the top and bottom edges, and subjected to a uniaxial in-plane compression along the vertical
direction by reducing the distance between the plates. Using image analysis
(see Materials and Methods) we computed the beams' tortuosity, defined as the ratio 
of the contour length of each beam to its end-to-end distance. 
Buckled beams have a tortuosity 
significantly above 1. Under a vertical compression of 4\%, only 
beams along the left domain wall show a tortuosity above 1.05,  consistent 
with a localized buckling response (Fig.~5b). However, under higher strains, the 
response of the lattice qualitatively changes. When beams have buckled along 
the entire domain wall, its response to further loading is no longer compression-dominated 
(Supplementary Figure S1). Future buckling events, which are triggered by 
coupling between torsional and compressional forces on beams due to the stiff 
hinges, no longer single out the left domain wall and 
happen uniformly throughout the sample (Fig.~5c and Supplementary Movie 2).

\begin{figure}
    \includegraphics{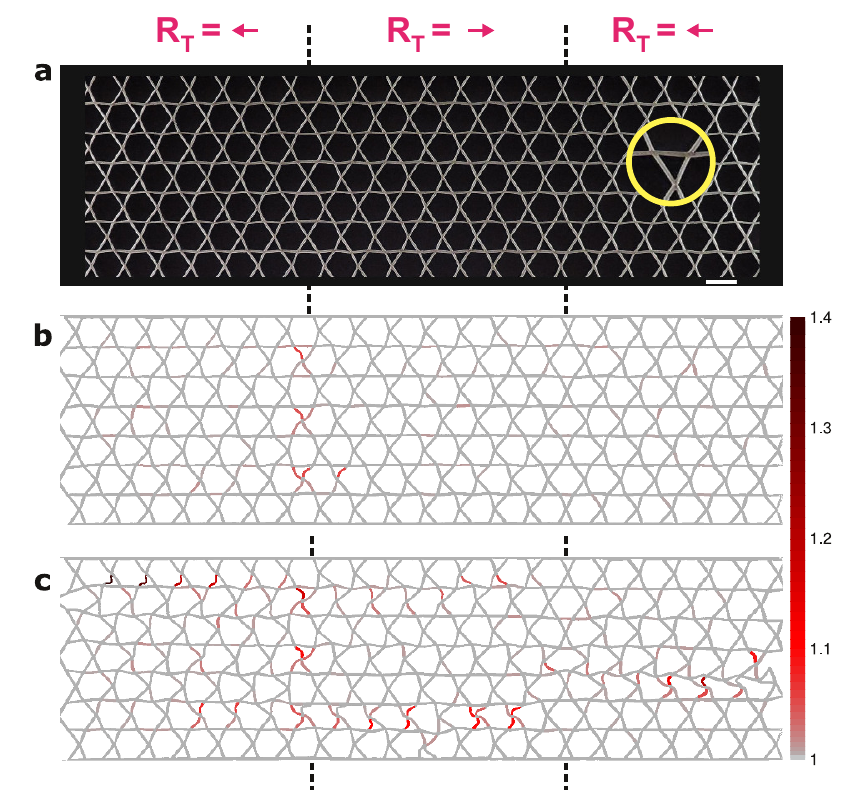}
    \caption{
        {\bfseries Buckling is robust under polarization-preserving changes of the unit cell.} 
        {\bfseries a,} A 2D foam cellular prototype, whose unit cell 
        maintains the topological polarization $\Rt$ even though its distortion 
        away from the regular kagome lattice is small (a zoom of the constituent 
        triangles is shown within the yellow circle). The 
        domain wall geometry is identical to that of the 3D sample, with the 
        left domain wall localizing states of self-stress. Scale bar, 2 cm. 
        {\bfseries b,} Response of the structure under a vertical compression 
        of 4\% with free left and right edges. The beams are coloured by the 
        tortuosity, the ratio of the initial length of the beam to the 
        end-to-end distance of the deformed segment (colour bar).
        {\bfseries c,} Response of the structure under 7\% compression, with 
        beams coloured by tortuosity using the same colour scale as in 
        {\bfseries b}.
    } \label{fig_robust}
\end{figure}

We have demonstrated that piling up localized states of self-stress in a small portion 
of an otherwise bending-dominated cellular metamaterial can induce a local 
propensity for buckling. Whereas this principle is of general applicability, 
our buckling regions exploit topological states of self stress~\cite
{Kane2014} which provide two advantages. First, they are indistinguishable from the rest of the structure in 
terms of node connectivity and material parameters, allowing
mechanical response to be locally modified without changing the
thermal, electromagnetic or optical properties. This 
feature could be useful for optomechanical~\cite{Eichenfield2009a} or 
thermomechanical~\cite{GudaVishnu2013} metamaterial design.  Second, 
the buckling regions are robust against structural perturbations, as long as the acoustic bulk gap 
of the underlying frame is maintained. This gap is a property of the 
unit cell geometry. Large deformations that close the gap 
could be induced through external actuation~\cite{Leung2007} or confinement, potentially allowing a tunable 
response from localized to extended buckling 
in the same sample, reminiscent of electrically tunable band gaps in 
topological insulators~\cite{Drummond2012}.

{\section{Materials and Methods}

\subsection{Deformed kagome lattice unit cell}
The deformed kagome lattices are obtained by decorating a 
regular hexagonal lattice, built from the primitive lattice vectors 
$\{\mathbf{a}_1=a\hat{x},\,\mathbf{a}_2=-(a/2)\hat{x}+(\sqrt{3}a/2)\hat{y}\}$ 
where $a$ is the lattice constant, with a three-point unit cell that results 
in triangles with equal sizes. The unit cells  
are described by a parametrization introduced in Ref.~\cite{Kane2014} 
which uses three numbers $(x_1,x_2,x_3)$ to quantify the distortion of lines 
of bonds away from a regular kagome lattice ($x_i=0$). 
The unit cell in Fig.~1a, which forms the basis for the 3D cellular 
structure, is reproduced by $(x_1,x_2,x_3)=(-0.1,0.06,0.06)$ for which the 
topological polarization is $\Rt = -\mathbf{a}_1$~\cite{Kane2014}. The unit 
cell for the design in Fig.~5a parametrized by $(-0.025,0.025,0.025)$ has the 
the same polarization. In all cases we describe the unit cell and polarization in
the outer region; the inner region between the domain walls rotates this unit cell
by $\pi$ which flips the polarization direction.
    
\subsection{Construction of the equilibrium/compatibility matrix}
Analysis of the linear response of a frame or a cellular material begins with 
the construction of the equilibrium matrix $\mathbf{A}$. Since it is more natural to relate
node displacements to beam length changes, we describe 
how to build the compatibility matrix $\mathbf{C} = \mathbf{A}^\text{T}$, 
relating point displacements $\mathbf{u}$ to extensions 
$\mathbf{\vare}$ via $\mathbf{C}\mathbf{u} = \mathbf{\vare}$. We construct
the compatibility matrix from the contributions of individual beams. 
Consider a single beam aligned to the $x$ axis connecting hinge 1 at $(0,0)$ to  hinge 2 
at $(L,0)$. There are six degrees of freedom potentially constrained by the 
beam: the positions $(x_i,y_i)$ at each node $i$ and the torsion angle 
$\theta_i$. Within Euler-Bernoulli beam theory, these values are sufficient to 
determine the shape of the beam along its length as well as the forces and 
torques at each end needed to maintain equilibrium~\cite{Logan2007}. Three independent 
combinations of forces and torques can be identified which are proportional to 
generalized strains experienced by the beam: pure tension $t \propto x_2-x_1$, pure shear $s \propto 
y_2-y_1-L(\theta_1+\theta_2)/2$, and pure bending torque $m \propto 
\theta_2-\theta_1$, as illustrated schematically in Fig.~3. In matrix form, 
this gives
\begin{equation}
    \mathbf{\vare} = 
        \begin{pmatrix}
            -1 & 0 & 0 & 1 & 0 & 0 \\
            0 & -1 & -L/2 & 0 & 1 & -L/2 \\
            0 & 0 & -1 & 0 & 0 & 1
        \end{pmatrix} \cdot \begin{pmatrix}x_1 \\ y_1 \\ \theta_1 \\ x_2 \\ 
        y_2 \\ \theta_2 \end{pmatrix} \equiv \mathbf{Cu}. \label{eqn_cmx}
\end{equation}
The forces and torque (generalized stresses) are obtained from the generalized strains 
$\mathbf{\vare}$ through the stiffness matrix $\mathbf{K}$ which depends on 
the Young's modulus $E$, the cross-sectional area  $A$, the area moment of 
inertia $I$ and the beam length $L$:
\begin{equation}
    \bs{\sigma} \equiv \begin{pmatrix}t \\ s \\ m\end{pmatrix}
        = \begin{pmatrix}
            EA/L & 0 & 0 \\
            0 & 12 EI/L^3 & 0 \\
            0 & 0 & EI/L
          \end{pmatrix} \cdot \mathbf{\vare} \equiv 
          \mathbf{K}\mathbf{\vare}.
\end{equation}
The compatibility matrix for a beam with arbitrary orientation is
obtained by projecting the displacement vectors at the end of each beam onto 
the axial and transverse directions using the 
appropriate rotation matrix which depends on the angle made by the beam with 
the $x$ axis. Each beam in an assembly provides three rows to the 
compatibility matrix, with additional columns set to zero for the degrees of 
freedom unassociated with that beam. 
In the simpler frame limit, each beam only resists axial 
extensions , and contributes  one row (the first row of the 
$\mathbf{C}$ matrix in Eq.~\ref{eqn_cmx}) to the compatibility matrix. 

Once the equilibrium matrix is constructed, the approximate states of 
self-stress of the periodic frame (Fig.~2) as well as the linear response of 
the cellular material under loads (Fig.~3) are obtained from its singular value 
decomposition following the methods of Ref.~\cite{Pellegrino1993}. More 
details of the computation are provided in the SI Text.

\subsection{Construction and characterization of 2D and 3D prototypes}
The 3D structures were printed by Materialise N.V. (Leuven, BE) through laser sintering of their proprietary thermoplastic polyurethane TPU 92A-1 (tensile strength 27 MPa, density 1.2 g/cm$^{3}$, Young's modulus  27 MPa).

The 2D structures were cut using a VersaLaser 3.5 laser cutter (Laser \& Sign Technology, New South Wales, AU) from 1.5 cm thick sheets of closed-cell cross-linked polyethylene foam EKI-1306 (EKI B.V., Nijmegen, NL; tensile strength 176 kPa, density 0.03 g/cm$^{3}$, Young's modulus 1.7 MPa).
A characteristic load-compression curve of a 2D sample is shown in Supplementary Figure S3.

\subsection{Image analysis of 2D experiment}
Images of the 2D cellular prototype (Fig.~5) were obtained using a 
Nikon CoolPix P340 camera and stored as 3000x4000 px 24-bit JPEG images. 
To quantitatively identify the buckled beams in the 2D prototype under 
confinement, we extracted the {\it tortuosity} $\tau$ of each beam, defined 
as the ratio of the length of the beam to the distance between its  
endpoints. Tortuosity was estimated from the sample images through a series of
morphological operations, as detailed in SI Text and Supplementary Figure S4.
Straight beams have $\tau = 1$ whereas beams that buckle 
under axial compression have $\tau > 1$. Unlike other measures of curvature, 
tortuosity distinguishes buckled beams from sheared beams (schematic, Fig.~3b) 
which have $\tau \gtrsim 1$.

\begin{acknowledgments}
We thank Denis Bartolo, Bryan Chen, Martin van Hecke, Charlie Kane, and Tom 
Lubensky for useful discussions, and Jeroen Mesman at the Leiden University Fine Mechanics  
Department for designing and building the compression stage. This work was funded by FOM 
and by the D-ITP consortium, a program of the Netherlands Organisation for 
Scientific Research (NWO) that is funded by the Dutch Ministry of Education, 
Culture and Science (OCW).
\end{acknowledgments}

\bibliography{refs}

\newpage
\setcounter{figure}{0}
\setcounter{equation}{0}
\setcounter{section}{0}

{\sffamily \Large {\bfseries Supplementary Information}}

\renewcommand{\theequation}{S\arabic{equation}}
\renewcommand{\thefigure}{S\arabic{figure}}

\section{SVD analysis of equilibrium matrix}
The linear response of the frame as well as the cellular metamaterial are 
obtained using the singular value decomposition (SVD) of the equilibrium 
matrix, as detailed in Ref.~\cite{Pellegrino1993} and summarized below. The SVD analysis 
simultaneously handles the spaces of node forces as well as beam stresses, and 
properly takes into account states of self-stress in the structure. 

The SVD of the equilibrium matrix $\mathbf{A}$ with $\nr$ rows and $\nc$ 
columns expresses it as the product of 
three matrices:
\begin{equation}
    \mathbf{A} = \mathbf{UVW^\text{T}},
\end{equation}
where $\mathbf{U} = [\mathbf{u}_1,...,\mathbf{u}_{\nr}]$ is a square matrix 
composed of $\nr$ independent 
column vectors $\mathbf{u}_i$, $\mathbf{W} = [\mathbf{w}_1,...,\mathbf{w}_{\nc}]$
is a square matrix of $\nc$ independent column vectors $\mathbf{w}_i$, and 
$\mathbf{V}$ is a matrix with $r$ non-negative values $v_{ii}$ on the leading diagonal 
and all other values zero, $r$ being the rank of $\mathbf{A}$.

The first $r$ columns $\mathbf{u}_i$ of $\mathbf{U}$ provide the finite-energy displacement 
modes of the structure, whereas the remaining $\nr-r$ column vectors are the 
zero modes. Similarly, the first $r$ and remaining $\nc-r$ column vectors 
of $\mathbf{W}$ provide the finite-energy bond tension configurations and the 
states of self-stress respectively.

For the frame in main text Fig.~2 under periodic boundary conditions, the 
count of topological states of self-stress predicts eight localized states of 
self-stress if the left domain wall were isolated, in addition to 
the two extended states of self-stress expected under periodic boundary 
conditions. However, since 
the separation between the left and right domain walls is finite, the SVD 
produces only two actual states of self-stress, and six 
approximate states of self-stress with large $v_{ii}$. The corresponding 
tension configurations $\tss{q}\equiv \mathbf{w}_{\nc-q}$ are the last eight column vectors of 
$\mathbf{W}$. For of these ($\mathbf{w}_{\nc-7}$, $\mathbf{w}_{\nc{}-6}$, $\mathbf{w}_{\nc-1}$, and
$\mathbf{w}_{\nc}$) have a significant overlap with $\uxx$ or $\uyy$. 
These are displayed in Fig.~2a--d respectively. 

The response of the cellular network in Fig.~3 of the main text is calculated for the finite 
block with free edges. Unlike the frame, this is highly overconstrained and 
has many generalized states of self-stress (now corresponding to mixtures of 
tensions, shears and torques that maintain equilibrium).  
An external load $\mathbf{l}$, 
corresponding to a force specification on each point, is supportable by the 
structure provided its overlap with the space of zero-energy motions is zero; 
i.e. 
\begin{equation}
    [\mathbf{u}_{r+1},...,\mathbf{u}_{\nr}]^\text{T} \mathbf{l} = \mathbf{0}.
\end{equation}
Since the cellular block only has three zero motions corresponding to the two 
translations and one rotation, any set of forces with no net force or torque 
on the structure, such as the force configuration shown in Fig.~3a, satisfies 
this condition. The generalized stresses $\bs{\sigma}$ (three per beam) are then given by
\begin{equation}
    \bs{\sigma} = \sum_{i=1}^r \frac{ \mathbf{u}_i\cdot 
    \mathbf{l}}{v_{ii}}\mathbf{w}_i + \mathbf{W}_{\nc-r} \mathbf{x},
\end{equation}
where the second term is an arbitrary combination of the states of self-stress 
given by the last $\nc-r$ columns of $\mathbf{W}$. The weights $\mathbf{x}$ are 
determined by requiring zero overlap of the generalized strains with the 
incompatible strains (states of self-stress), which leads to the 
equation
\begin{equation}
    \mathbf{W}^\text{T}_{\nc-r} \mathbf{K}^{-1} \mathbf{W}_{\nc-r} \mathbf{x} 
    = -\mathbf{W}^\text{T}_{\nc-r} \mathbf{K}^{-1}\sum_{i=1}^r \frac{ \mathbf{u}_i\cdot 
    \mathbf{l}}{v_{ii}}\mathbf{w}_i.
\end{equation}
which can be solved for $\mathbf{x}$ to get the complete generalized stresses 
(three per beam, plotted separately in main text Figs.~3a--c).

\section{Sequential removal of beams}
\begin{figure*}
    \centering
    \includegraphics[width=80mm]{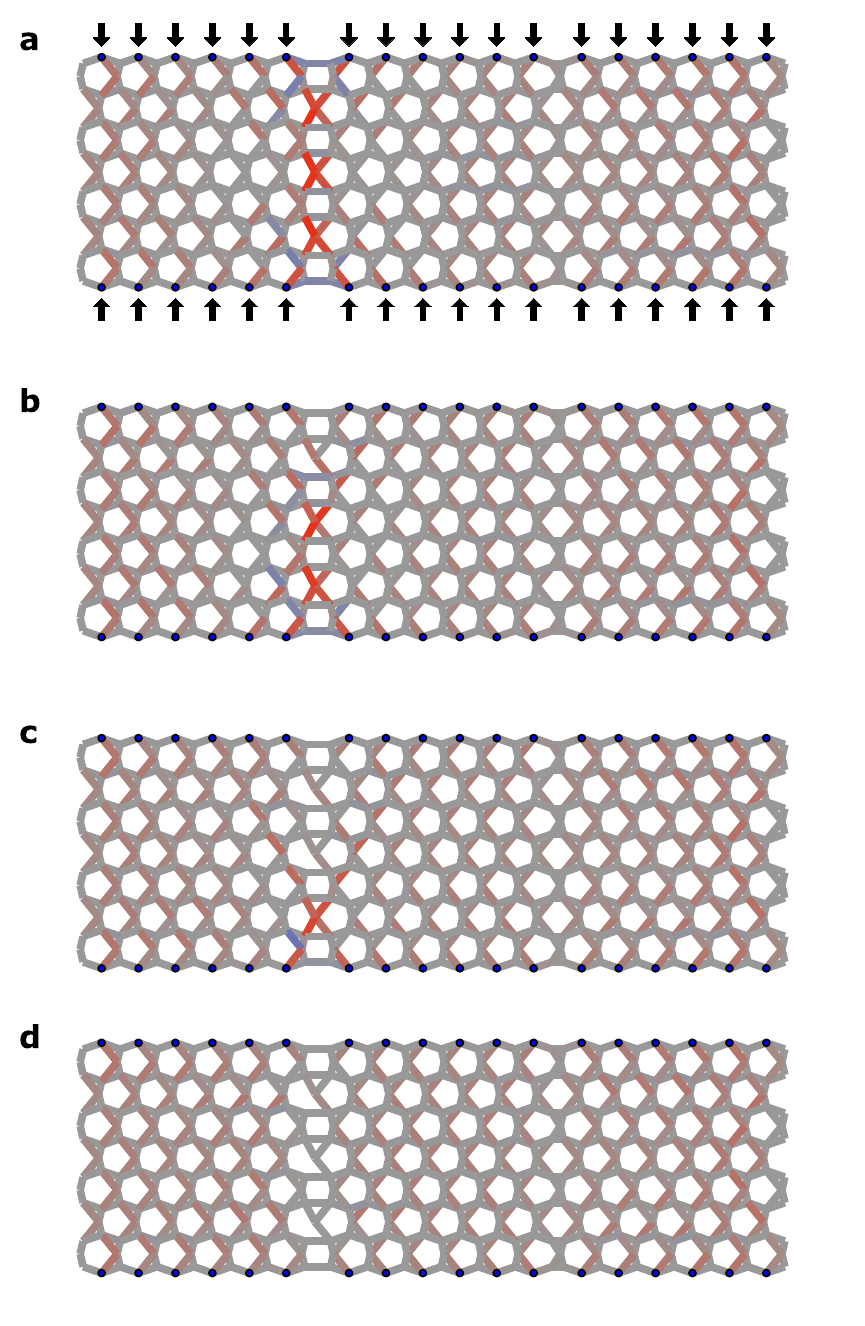}
    \caption{{\bfseries Sequential removal of beams to recreate the effect of 
    buckling in the linear response.}
    {\bfseries a,} Axial compression of the cellular material shown in Fig.~3a under 
    identical loading, but with beams coloured by the propensity for buckling, 
    $-t_iL_i^2/F$. The beam with the highest value of this quantity will be 
    the first to buckle as the force $F$ on the boundary points is ramped up. 
    {\bfseries b--d,} Result of sequentially removing the beam with the 
    highest propensity for buckling and recalculating the compressions in 
    the remaining beams under the same axial loading.\label{fig_si_sequential}} 
\end{figure*}

Fig.~3 (main text) showed that the axial compressions in the cellular material 
under uniaxial loading were concentrated along the left domain wall, matching 
the expectation from the analysis of the frame with similar beam geometry. 
However, every time a beam buckles, its axial load-bearing ability is lost. 
This fundamentally changes the load-bearing states of the underlying frame, 
and there is no guarantee that the beams with the highest compression continue to be along 
the left domain wall. However, the number of topological states of self-stress 
localized to the left domain wall in the underlying periodic frame grows 
linearly with the length of the domain wall~\cite{Kane2014}, suggesting that 
the load-bearing ability of the left domain wall may not vanish completely 
with just a single buckling event. Here, we numerically show using a sequential 
analysis of the finite cellular frame that the left domain wall can 
support as many buckling events as there are repeating units along the $y$ 
direction.

To recap, Fig.~3a (main text) shows the compressions in a finite cellular block, obtained 
by tiling the same pattern three times along the $y$ direction, under forces 
applied to undercoordinated points at the top and bottom edge.
Since the threshold for buckling under compression scales as $1/L_i^2$ for  
beam $i$ of length $L_i$, we expect that the beam with the largest value of 
$-t_iL_i^2$  will buckle first as the compressive force $F$ is increased. We examine the influence 
of the buckling event on the compression response by removing this beam from 
the cellular structure and recalculating the axial tensions $\mathbf{t}$ 
without changing the forces on the boundary points. This process can be 
repeated, sequentially removing the beam with the largest value of $-t_iL_i^2$ 
after the linear response of the cellular structure has been recalculated. The 
result is show in Fig.~\ref{fig_si_sequential}, in which beams are coloured by 
$-t_iL_i^2/F$, i.e. the beam with the highest intensity in each iteration is 
removed in the next. For the first three iterations 
of the sequential process, the beam chosen for removal lies along the left 
domain wall (Figs.~\ref{fig_si_sequential}a--c), showing that the multiplicity 
of load-bearing states localized to the left domain wall is sufficient for 
at least one buckling event to occur for each repeating unit of the domain 
wall. 

Furthermore, after each repeating unit along the $y$ direction has experienced 
buckling, the beams in the rest of the lattice experience much lower 
compressions relative to the initial compressions along the left domain wall 
(Fig.~\ref{fig_si_sequential}d), signifying that the lattice remains entirely 
bending/shear-dominated away from the domain wall.

By performing simulations on different system sizes, we confirmed that upon increasing the sample size 
along the $y$ direction, the number of buckling events localized to the left domain wall increases 
proportionally.

\begin{figure*}[t]
    \centering
    \includegraphics[width=120mm]{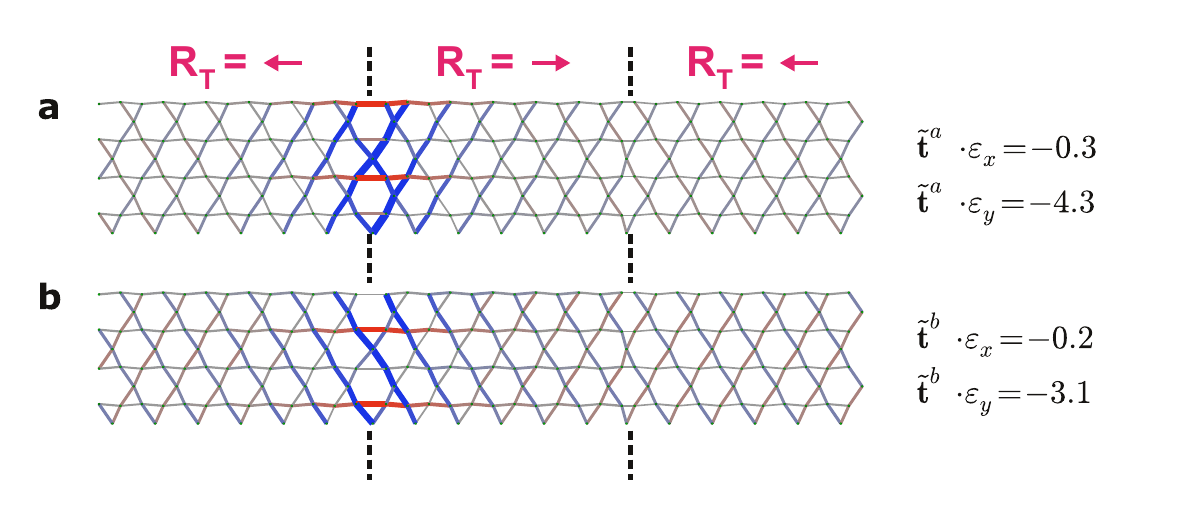}
    \caption{{\bfseries Localized states of self-stress for frame
    corresponding to design of Fig.~5a (main text).}
    The numerically-obtained states of self-stress for a frame with the same 
    domain wall geometry as in Fig.~2, but with the unit cell 
    parametrized by $(x_1,x_2,x_3)=(-0.025,0.025,0.025)$ which 
    has a smaller distortion away from the regular kagome lattice and is used 
    in Fig.~5a. The left 
    domain wall still has a net outflux of 
    the topological polarization and localizes states of self-stress, which 
    have a significant overlap with the affine bond extensions $\bs\ex$ and 
    $\bs\ey$ associated with uniform strains along $x$ and $y$ respectively.} \label{fig_si_selfstress}
\end{figure*}

\section{Load-compression curve of a 2D sample}

\begin{figure*}
	\centering
	\includegraphics[width=80mm]{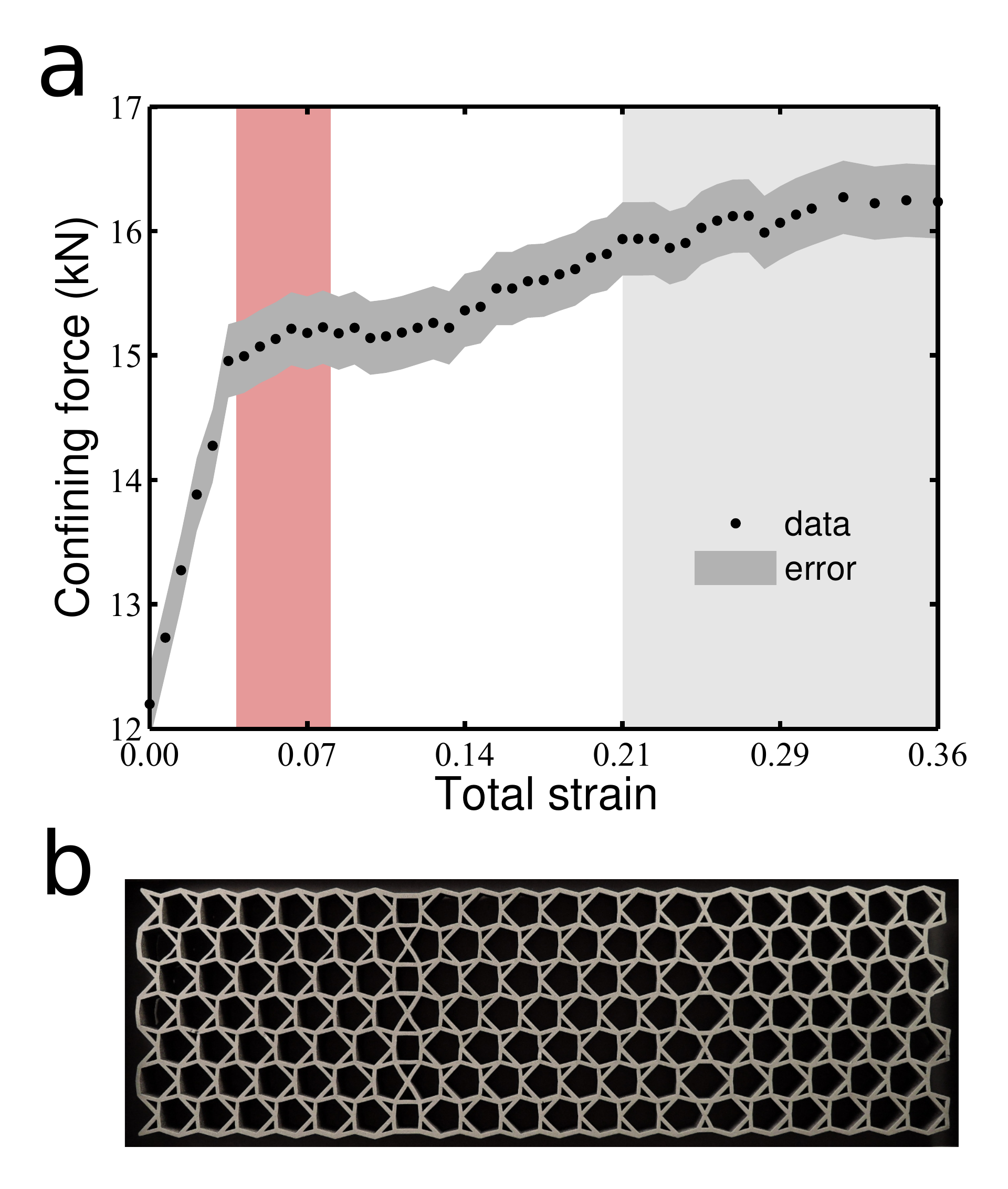}
	\caption{{\bfseries Measurement of the total confining force as a
            function of imposed strain for a 2D cellular prototype sample
            (Fig.~5).}
          {\bfseries a,} The measured data (black circles) and an
          estimate of the measurement error (dark grey contour) are shown. The
          red region indicates the strain regime in which the beams within the
          topologically rigid domain wall buckle. In the light grey region, the
          sample response is dominated by out-of-plane buckling of the entire
          sample rather than by in-plane deformations of the cellular
          structure. The error was estimated at 0.2 kN based on uncertainties in
          the compression and force due to the rudimentary measurement
          apparatus.
          {\bfseries b,} 2D cellular prototype for which the
          load-compression curve was measured. Its unit cell in the outer region
          is parametrized by
          $(x_1,x_2,x_3)=(-0.085,0.085,0.085)$.}\label{fig:forcecurve}
\end{figure*}

A rudimentary measurement (supplementary figure~\ref{fig:forcecurve}) of the
load-compression curve of a 2D cellular prototype has been performed to assess
the effect of the localized buckling events on the mechanical response. The
prototype, shown in Fig.~\ref{fig:forcecurve}b, uses a unit cell characterised
by $(x_1,x_2,x_3)=(-0.085,0.085,0.085)$ in the parametrization of
Ref.~\cite{Kane2014}. This is geometrically similar to the unit cell used for
the 3D sample in the main text and has the same polarization. However, we use
more unit cells in the 2D sample to reduce edge effects on the measurement. The
overall geometry is similar to that of the finite 2D lattice used in the linear response
calculations of main text Fig.~3.

 The 2D cellular foam sample (460 mm x 150 mm x 15 mm; beam widths 1.5--2.5 mm; see Materials and Methods for more details) was positioned with its narrow
edge against a supporting acrylic plate and scale (Kern PCB 10000-1). The
sample was confined from above by a second acrylic
plate. No lateral confinement was provided.
The distance between top and bottom plates was incrementally decreased
from 140 to 90 mm using a
laboratory jack. At each confinement step the system
was allowed to relax and the equilibrium force on the sample edge was recorded from the scale,
resulting in a quasistatic measurement of the sample's load-compression curve. A
small upward drift of ca. 0.5 kN was recorded in the scale's force response
due to long relaxation times. This drift (assumed to be linear in the
confinement) has been subtracted from the recorded data. The complete
measurement is shown in figure~\ref{fig:forcecurve}a. Buckling of the beams along the domain wall
leaves a clear signature in the form of a sudden softening of the response (region highlighted
in red). Prior to buckling, the main resistance to confinement was provided by the
compressed beams along the domain wall since the rest of the beams could easily bend to
accommodate the confinement. Buckling effectively removes the compressional stiffness of
the beams along the domain wall. Further confinement is only resisted by additional bending
of all beams, which leads to a significantly softer response.

A linear load-compression followed by a plateau is typical of elastomeric cellular
materials~\cite{Gibson1989}. Such materials also tend to have a secondary
stiffening regime due to densification when adjacent beams contact each other
along their lengths. In our measurement, however, the entire sample buckled in
the lateral direction before this regime could be reached.

\section{Details of 2D image analysis}
In the Materials and Methods section of the main text, we outlined the 
quantitative analysis of the 2D experimental data in Fig.~5 (main text). Here, 
we provide more details of the image processing required to measure the 
tortuosity of each beam in a cellular sample under compression. The tortuosity 
is defined as the ratio of the length of each beam to its end-to-end distance. 

The analysis starts with a raw image such as in Fig.~5a (main text), a 
top-down view of the sample whose top surface is selectively illuminated so 
that the beams making up the cellular material sample are visible against a 
dark background. Fig.~\ref{fig:demo}a shows a subset of a different sample 
with similar beam lengths and widths, on which we illustrate the image 
processing steps.

\subsection{Isolating beam sections}
\label{subsubsec:beams}			
                                
To quantify the tortuosity of the lattice beams, the neutral axis (center line) of each beam must be isolated. The image is first binarized through 
intensity thresholding (figure~\ref{fig:demo}b): pixels with an 
intensity value above/below a certain threshold are set to 1/0. This 
separates the structure (1) from the background (0).
        
The finite-width beams are then reduced to 1-pixel-wide center lines by finding the 
morphological skeleton of the binarized image. To do this, we perform a 
repeated sequence of binary morphological operations: skeletonizing 
and spur removal. The former consists of removing edge {\tt `1'} pixels without 
breaking up contiguously connected regions. The latter 
is a pruning method that removes protrusions under a certain 
threshold length; it reduces spurious features due to uneven 
lighting and resolution limitations. The resulting skeleton is shown in figure~\ref
{fig:demo}c.

Next, we break up the skeleton into the neutral axes of individual beams.
Note that the topology of the frame, with four beams coming together at each 
joint, is not reproduced in the skeleton, due to the finite size of the beam joints. Instead, the skeleton has short segments connecting 
junctions of pairs of beams. To remove these segments, we
identify the branch points 
of the skeleton (pixels with more than two neighbors) and
clear a circular area around them. The size of the segments and of the cleared disk is of the same order as the original beam width. The resulting image, consisting of the  beams' approximate 
neutral axes, is shown in figure~\ref{fig:demo}.

Once the beam axes have been separated, two steps are implemented 
to improve the signal to noise ratio. First of all, beams near the 
edges of the sample are ignored. They are not always separated 
correctly from their neighboring beams, since they have lower 
connectivity than beams in the bulk of the lattice; this can lead to 
overestimation of the tortuosity. Secondly, since our lattices are 
fairly regular, the beams have a restricted size. Beam elements 
above and below these length thresholds are artifacts, and we ignore 
them. Figure~\ref{fig:demo}e illustrates the processed image data 
after these final steps.

\begin{figure*}
\centering
\includegraphics[width=140mm]{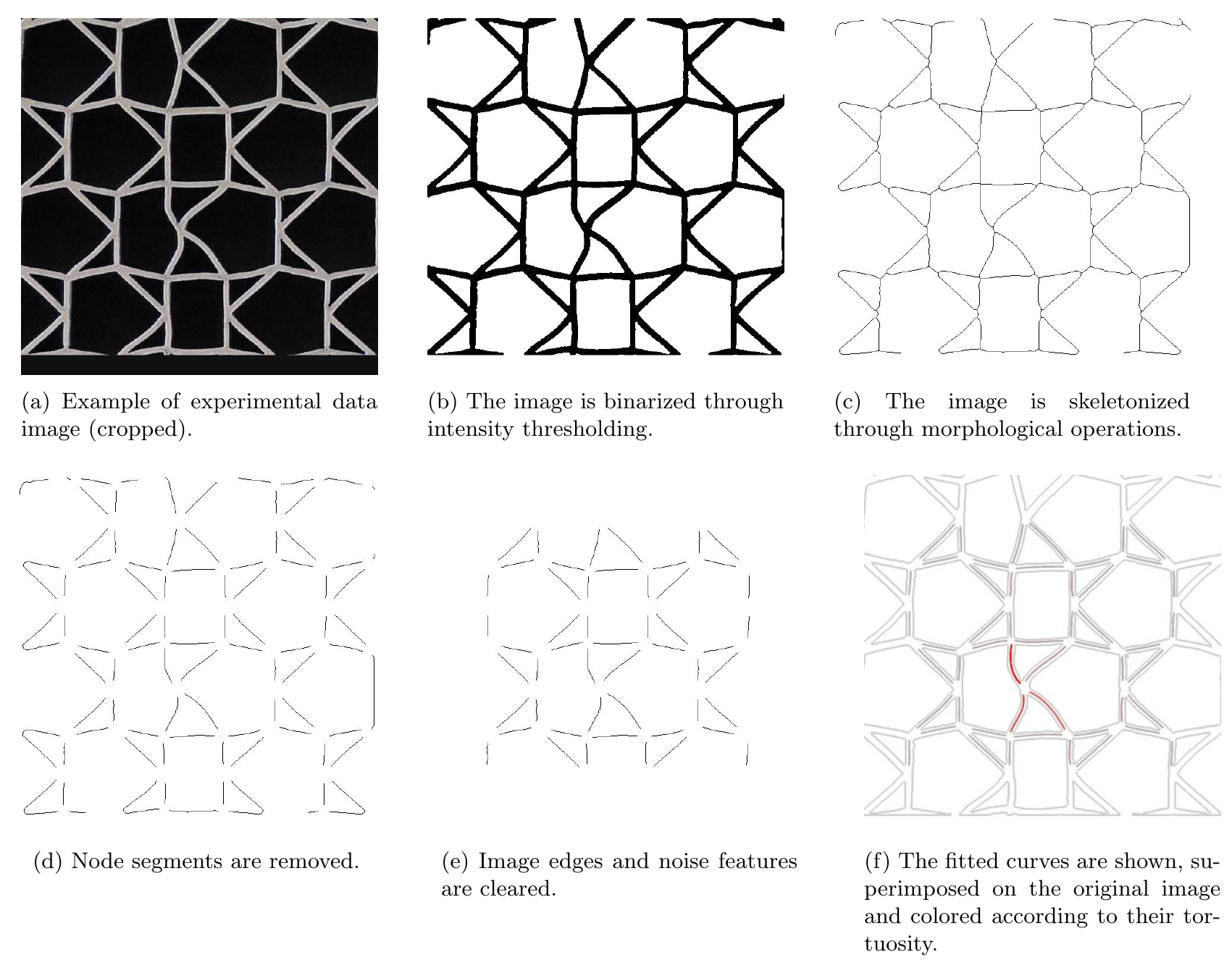}
\caption{Demonstration of steps in the image analysis process.}\label{fig:demo}
\end{figure*}
            
\subsection{Computing the tortuosity of each neutral axis}
\label{subsubsec:tortuosity}

After the beam axes have been isolated, connected components (sets of contiguous {\tt `1'} pixels) are found and labeled. Each 
connected component corresponds to an individual beam axis. Numerical data 
can be extracted from it by analyzing its pixels' coordinates, 
$\vec{r_p}=(x_p,y_p)$. These coordinates are fitted with polynomials $\left(x\left(l\right),y(l)\right)$ parametrized 
by the normalized path length along the beam, $l \in [0,1]$: 		 
\begin{align}\label{eq:fit2}
\begin{pmatrix}x(l)\\y(l)\end{pmatrix} = \sum\limits_{i = 0}^{3} \begin{pmatrix}a_i\\b_i\end{pmatrix} l^i
\end{align}
Cubic polynomials in $l$ successfully capture the shape of the neutral axis 
of each beam, while smoothing out pixel noise which leads to small-wavelength 
features in the binary skeleton. Fig.~\ref{fig:demo}f displays the smoothed contours superimposed on 
the original image, showing that the shapes of the individual beam sections 
are faithfully reproduced by the parametric form of Eq.~\ref{eq:fit2}.
        
The tortuosity $\tau$ of the beam axes is the arc length $A$ along the line divided by the distance $D$ between the endpoints. Both quantities can be extracted from the fitted model of the beam axes, $\vec{r}(l)=(x(l),y(l))$:
\begin{align}\label{eq:tortuosity}
A &= \int_0^1 \! \sqrt{ \left(\frac{dx}{dl}\right)^2 + \left(\frac{dy}{dl}\right)^2 } \, dl \\
D &= |\vec{r}(1) - \vec{r}(0)|
\end{align}
Once the tortuosities of the beam segments have been computed, pixels in each 
segment of the binary image 
(Figure~\ref{fig:demo}b) are coloured according to the corresponding 
tortuosity to obtain images such as Figs.~5b--c of the main text.


\clearpage
\begin{center}
\includegraphics[width=87mm]{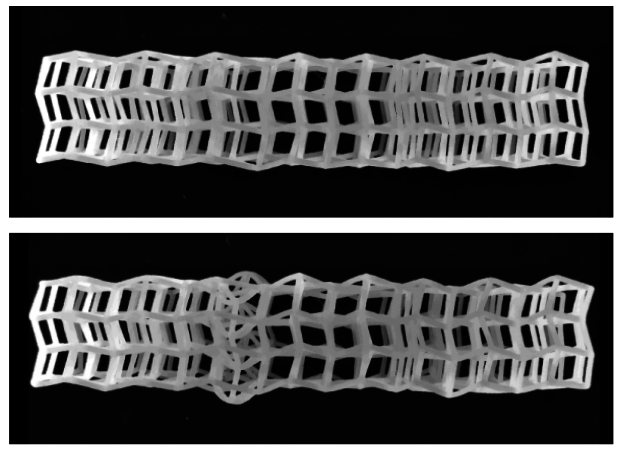}
\end{center}
{\bfseries Supplementary Movie 1:} View of the 3D layered cellular metamaterial prototype as
it is compressed into the plane of the figure*. A subset of beams singled out by 
topological states of self-stress in the underlying frame buckles out of the 
layer producing a distinctive visual signature.

\begin{center}
\includegraphics[width=87mm]{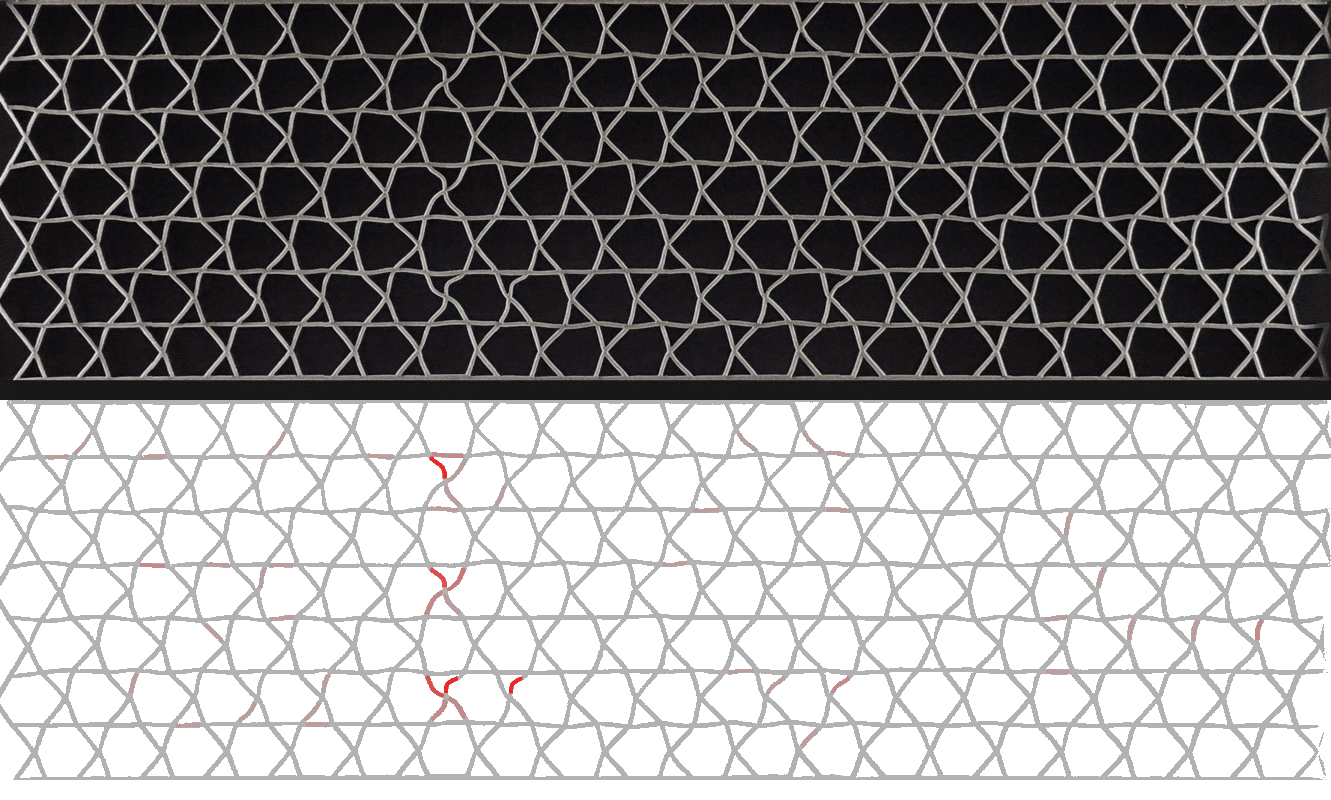}
\end{center}
{\bfseries Supplementary Movie 2:} View of the 2D cellular metamaterial prototype as
it is compressed vertically. Along with each raw image, a reconstructed image with beams coloured
by their tortuosity (as in main text Fig.~5) is also shown.

\end{document}